\documentclass[onecolumn]{aa}  
\usepackage{graphicx}
\usepackage[varg]{txfonts}
\usepackage{hyperref}
\usepackage{CJK}
\usepackage{booktabs}  
\usepackage{multirow}

\titlerunning{A Plateau-like EUV Late-phase Flare Driven by a MFR}
\authorrunning{Chen et al.}

\begin{document}

\begin{CJK*}{UTF8}{gbsn}
                
\title{An atypical plateau-like extreme-ultraviolet late-phase solar flare driven by the nonradial eruption of a magnetic flux rope}
                
\author{Yuehong~Chen ({\CJKfamily{gbsn}陈悦虹})\inst{1} \and Yu~Dai ({\CJKfamily{gbsn}戴煜})\inst{1,2} \and Mingde~Ding ({\CJKfamily{gbsn}丁明德})\inst{1,2}}
                
\institute{School of Astronomy and Space Science, Nanjing University, Nanjing 210023, People's Republic of China\\
\email{ydai@nju.edu.cn} 
\and Key Laboratory of Modern Astronomy and Astrophysics (Nanjing University), Ministry of Education, Nanjing 210023, People's Republic of China}
                
\date{}

\abstract{Recent observations in extreme-ultraviolet (EUV) wavelengths reveal an EUV late phase in some solar flares that is characterized by a second peak in the warm coronal emissions ($\sim$3 MK) occurring several tens of minutes to a few hours after the corresponding main flare peak.}
{Our  aim is  to clarify the physical origin of an atypical plateau-like EUV late phase in an X1.8-class solar flare occurring on 2011 September 7 from active region (AR) 11283.}
{We mainly took advantage of observations with the three instruments on board the \emph{Solar Dynamics Observatory} (\emph{SDO}). We first characterized the plateau-like late phase using EUV Variability Experiment (EVE) full-disk integrated irradiance observations and Atmospheric Imaging Assembly (AIA) spatially resolved imaging observations. Then we performed a nonlinear force-free-field (NLFFF) extrapolation of the AR magnetic fields based on the photospheric vector magnetogram with the  Helioseismic and Magnetic Imager (HMI), from which a filament-hosting magnetic flux rope (MFR) is revealed. The eruption of the MFR is tracked both in the plane of the sky (POS) and along the line of sight (LOS) through visual inspection and spectral fitting, respectively. Finally, we carried out differential emission measure (DEM) analysis to explore the thermodynamics of the late-phase loops.}
{The MFR shows a nonradial eruption from a fan-spine magnetic structure. The eruption of the MFR and its interaction with overlying arcades invoke multiple magnetic reconnections that are responsible for the production of different groups of late-phase loops.  Afterward, the late-phase loops enter a long-lasting cooling stage, appearing sequentially in AIA passbands of decreasing response temperatures. Due to their different lengths, the different groups of late-phase loops cool down at different  rates, which makes their warm coronal emission peaks temporally separated from each other. Combining the emissions from all late-phase loops together, an elongated plateau-like late phase is formed.}
{} 
                
                
\maketitle

\end{CJK*}

        \section{Introduction} \label{sec1}
        Solar flares are one of the most energetic phenomena in the solar atmosphere, whose observation history can date back to one and a half centuries ago \citep{carrington1859}. The energy powering solar flares originates from coronal magnetic fields. Through the magnetic reconnection process \citep{priest1992}, the free magnetic energy stored in the corona is rapidly converted into plasma heating and particle acceleration, which are further transferred to electromagnetic radiation at all wavelengths from radio to $\gamma$-rays \citep{benz2017}. Together with a bulk mass motion that is manifested as a coronal mass ejection (CME), these energy outputs may cause strong perturbations in the space weather and produce significant geomagnetic effects \citep{schwenn2006}.
        
        According to the standard two-ribbon solar flare model, as known as the CSHKP model \citep{carmichael1964,sturrock1966,hirayama1974,kopp1976}, a flare is driven by the eruption of a magnetic flux rope (MFR). Magnetic reconnection takes place between the legs of magnetic field lines stretched out by the erupting MFR, with the released energy transported in the forms of heat flux and/or nonthermal electrons down to the chromosphere, producing a pair of conjugate flare ribbons. Meanwhile, an evaporative flow is driven upward from the chromosphere to fill the reconnected field lines, brightening up flare loops as typically seen in soft X-ray (SXR) and extreme-ultraviolet (EUV) filtergrams.
        
        In the framework of the CSHKP model, flare emissions in SXR and EUV generally exhibit an impulsive increase followed by a monotonic gradual decay \citep{hudson2011a}, indicating a heating--cooling process in the flare loops with mass cycling between the chromosphere and corona. By using observations with the EUV Variability Experiment \citep[EVE,][]{woods2012} on board the \emph{Solar Dynamics Observatory} \citep[\emph{SDO},][]{pesnell2012}, \citet{woods2011} have recently discovered a second peak of the warm coronal emissions (e.g., the \ion{Fe}{XVI} 335~{\AA} emission at $\sim$3 MK) in some solar flares, which occurs several tens of minutes to a few hours after the corresponding main flare peak, and hence is termed as the  ``EUV late phase.'' In addition to the temporal delay, imaging observations further reveal that the late-phase emission comes from a set of higher and longer flare loops other than the main flare loops \citep{woods2011,hock2012,sun2013,dai2013,liu2013,liu2015,masson2017,dai2018a,zhou2019,wang2020}.
        
        It has been shown that EUV late-phase flares always arise from active regions (ARs) of multipolar configurations \citep{woods2011,li2014a,masson2017,dai2018a}, which often harbor complex magnetic topologies, such as fan-spine structures \citep{lau1990,priest1996,parnell1996} and/or quasi-separatrix layers \citep[QSLs,][] {priest1992,priest1995}. The eruption of a MFR in such magnetic environments may drive multiple magnetic reconnections at different locations and on different scales \citep[e.g.,][]{masson2009,aulanier2019}, hence facilitating the production of flare loops in distinct lengths \citep{hock2012,liu2013,sun2013,masson2017,dai2018a}. Nevertheless, where and how   the multiple reconnections are involved in an EUV late-phase flare are still not well understood.
        
        Extreme-UV late-phase emission actually reflects a hydrodynamic response of the late-phase loops to reconnection heating. Based on its later occurrence, the EUV late phase was at first explained by a secondary heating temporally separated from the main flare heating \citep{woods2011,hock2012,dai2013,zhou2019,wang2020}. If this is the case, the delayed heating should be much less energetic than the main flare heating, giving rise to late-phase loops just heated to an intermediate temperature. In an alternative scenario, however, both the main flare loops and the late-phase loops are supposed to be heated nearly simultaneously, while the separation of the late-phase peak from the main flare peak is simply due to a long-lasting cooling process in the much longer late-phase loops \citep{liu2013,li2014a,masson2017,dai2018a}. It was also pointed out that both mechanisms are working in some cases \citep{sun2013,zhong2021}.
        
        Using the zero-dimensional (0D) enthalpy-based thermal evolution of loops \citep[EBTEL,][]{klimchuk2008,cargill2012,barnes2016} model, \citet{dai2018b} numerically studied the production of EUV late-phase flares under the above two mechanisms. It was found that  around the warm coronal late-phase peak, the late-phase loops may experience different cooling stages for the two different scenarios: conductive cooling   for the additional heating scenario and radiative cooling   for the long-lasting cooling scenario. This  difference will cause distinctions in the shape of the synthetic light curves, which may serve as a diagnostic tool to observationally differentiate between the two mechanisms.
        
        In general, an EUV late-phase flare exhibits a well-resolved late-phase peak. However, in a comparative study of homologous late-phase flares from an AR, \citet{zhong2021} found that the late phase in one flare event shows an atypical emission plateau that lasts for almost one hour. In this paper we present observations of this atypical event. An in-depth analysis reveals that the plateau-like late-phase emission originates from the long-lasting cooling of different groups of late-phase loops, which are produced in multiple magnetic reconnections driven by the nonradial eruption of a MFR. In Sect. \ref{sec2} we characterize the plateau-like EUV late phase. In Sect. \ref{sec3} we track the eruption process of the MFR. The multiple magnetic reconnections driven by the erupting MFR are depicted in Sect. \ref{sec4}, and the long-lasting cooling process of the late-phase loops is established in Sect. \ref{sec5}. Finally, we summarize the results and draw our conclusions in Sect. \ref{sec6}.
        
        \section{Plateau-like EUV late phase}\label{sec2}
\begin{figure*}
                \centering
                \includegraphics[width=12cm]{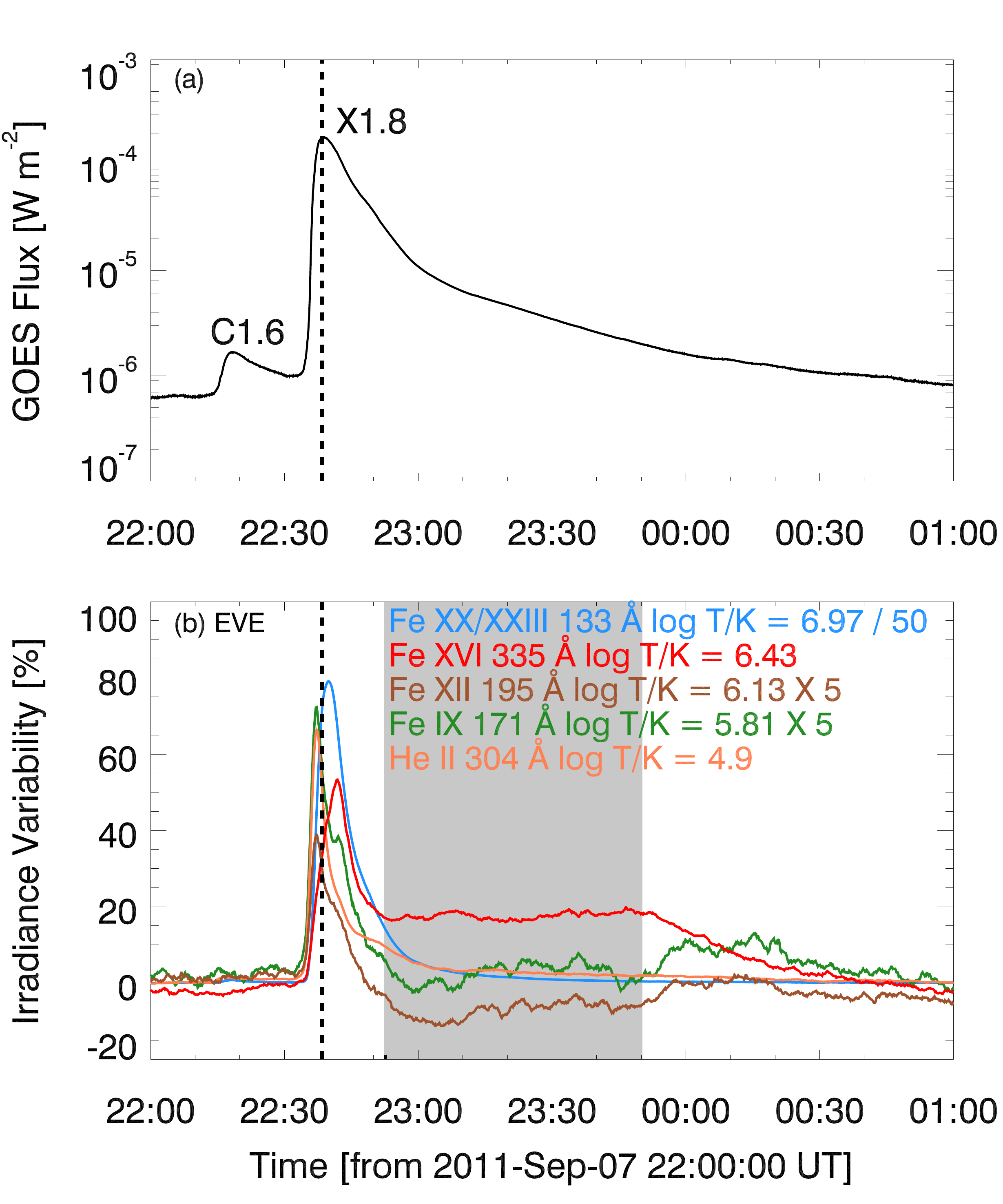}
                \caption{Time profiles of the \emph{GOES} 1--8 {\AA} flux (a) and   irradiance variability in several EVE spectral lines (b) for the 2011 September 7 X1.8-class flare. The vertical dashed line denotes the peak time of the SXR flux, and the shaded region highlights the duration of the emission plateau seen in EVE 335 {\AA}.  When deriving the EVE irradiance variability, a background level is calculated by averaging the absolute irradiance over a quiescent period of 21:30--22:00 UT\@. In addition,  a two-minute smoothing boxcar is applied to the EVE data points to enhance the signal-to-noise ratio.}  
                \label{fig1}
        \end{figure*}

        The flare under study occurred on 2011 September 7 from  AR NOAA 11283, a flare-productive AR that has been extensively studied in the literature \citep[e.g.,][]{jiang2013,jiang2014,liu2014,xu2014,romano2015,ruan2015,zhang2015,dai2018a,zhong2021}. According to the \emph{GOES} 1--8 {\AA} light curve shown in Fig. \ref{fig1}a, the flare starts at 22:32 UT, promptly reaches its peak at 22:38 UT, and ends at 22:44 UT, registered as an X1.8-class flare. As a possible precursor of this major flare, there is another C1.6-class flare that takes place about 20 minutes earlier from the same AR.    
        
        To trace the flare emissions at different temperatures, we plot in Fig. \ref{fig1}b the time profiles of the irradiance variability in several EVE spectral lines. EVE measures full-disk integrated EUV irradiance between 1 {\AA} and 1050 {\AA} with a time cadence of 10 s and a spectral resolution of 1 {\AA}, from which the irradiance of some ``isolated'' lines can be derived by spectral integration over specified wavelength windows. In this work we focus on observations from the Multiple EUV Grating Spectrograph A (MEGS-A) channel of EVE, which enables a nearly 100\% duty cycle over a spectral range of 65--370~{\AA}. As shown in the figure, the cold chromospheric \ion{He}{II} 304 {\AA} ($\log T\sim4.9$) emission exhibits an impulsive enhancement and reaches its maximum before the \emph{GOES} SXR peak, while the hot coronal \ion{Fe}{XX/XXIII} 133~{\AA} ($\log T\sim7.0$) emission closely resembles the SXR evolution and peaks slightly later. After the flare peak, the cool coronal \ion{Fe}{XII} 195 {\AA}  ($\log T\sim6.1$) and \ion{Fe}{IX} 171~{\AA} ($\log T\sim5.8$) emissions quickly turn to a dimming, probably due to mass drainage related to the CME lift-off \citep{reinard2008}. All the above patterns conform to the typical emission characteristics  of an eruptive flare \citep[cf.][]{woods2011}. Interestingly, the warm coronal \ion{Fe}{XVI} 335 {\AA} ($\log T\sim6.4$) emission reveals an elevated plateau following the main flare peak, which keeps a rather flat level, $\sim$35\% of the main peak emission, for almost one hour (highlighted by the gray shaded region). Superimposed on this plateau, there are also some small emission bumps, which seem to leave imprints in the cooler emissions (e.g., EVE 171~{\AA}) in the later phase.
        
\begin{figure*}
                \centering
                \includegraphics[width=12cm]{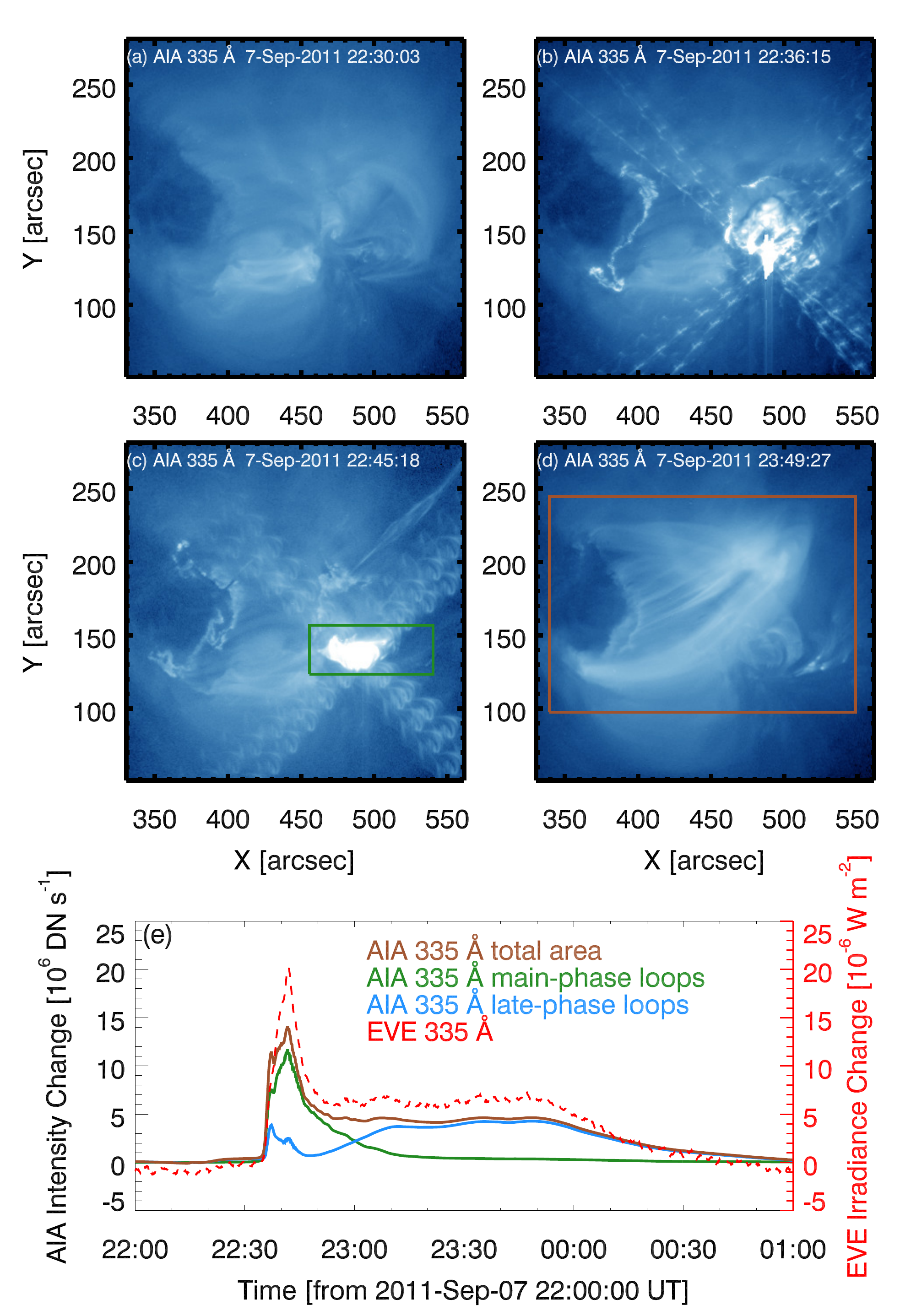}
                \caption{Snapshots of  flare evolution observed in the AIA 335~{\AA} passband (a--d), and background-subtracted AIA 335~{\AA} intensity profiles summed over specified regions (e). Shown are the whole flare-hosting AR (enclosed by the brown box in panel d), the main flare region (outlined by the green box in panel c), and the late-phase loop region (outside the main flare region, but within the AR). For comparison, the background-subtracted line irradiance in EVE 335 {\AA} is also overplotted as the dashed line in panel e.}\label{fig2}
        \end{figure*}

        To further determine the spatial distribution of the flare emissions, we resort to spatially resolved observations with the Atmospheric Imaging Assembly \citep[AIA,][]{lemen2012} on board \emph{SDO}. AIA provides simultaneous full-disk images of the transition region and corona in ten passbands with a pixel size of 0.6{\arcsec} and a temporal resolution of 12 s or 24 s. The online Animation 1 and Fig. \ref{fig2} demonstrate the flare evolution observed in AIA 335 {\AA}, a passband with a   temperature coverage similar to that of the EVE 335 {\AA} line. The flare starts with the destabilization and eruption of an east--west (EW) oriented filament that is initially located at the western part of the AR (Fig. \ref{fig2}a). The filament eruption brightens up multiple flare ribbons, including a quasi-circular ribbon (Rc) surrounding the main eruption site, a less discernible inner one (Ri) parallel to the southern segment of Rc, and an elongated remote one (Rr) over 50{\arcsec} to the east (Fig. \ref{fig2}b). Connecting these flare ribbons, different sets of post-flare loops are observed at different times. Shortly after the eruption of the filament, intensively brightening flare loops first appear within the main flare region, being organized along the location of the pre-eruption filament (Fig. \ref{fig2}c). As the main flare loops gradually fade off, several other sets of flare loops that link to the remote ribbon brighten up successively outside the main flare region (Fig. \ref{fig2}d). Compared to the main flare loops, these loops are much longer and more diffuse, and their overall brightening lasts for a significantly longer time. Such loop properties are consistent with the definition of late-phase loops by \citet{woods2011}.  
        
        Figure \ref{fig2}e displays the background-subtracted AIA 335 {\AA} intensity profiles summed over specified regions. For comparison, we  overplotted the background-subtracted line irradiance in EVE 335 {\AA}. First, the AIA 335~{\AA} profile of the whole flare-hosting AR (enclosed within the brown box in Fig. \ref{fig2}d) also exhibits a plateau following the main peak; its overall shape is very similar to that of the EVE 335~{\AA} curve. The similarity between the two profiles suggests that the variation in the EVE full-disk integrated irradiance predominantly comes from the AR\@. Second, when dividing the AR into subregions, it is seen that the main flare peak is mainly contributed by the emission from the main flare region (outlined by the green box in Fig. \ref{fig2}c), while the long-duration plateau after the main peak is dominated by the emission from the late-phase loops (region outside the main flare region, but still within the AR). This further validates the identification of an EUV late-phase flare for this event, although the late phase here shows an atypical emission plateau rather than a well-resolved peak in most cases.
        
        \section{Eruption of a filament-MFR system} \label{sec3}
        \subsection{Identification of a filament-hosting MFR}\label{sec31}
\begin{figure*}
                \centering
                \includegraphics[width=14cm]{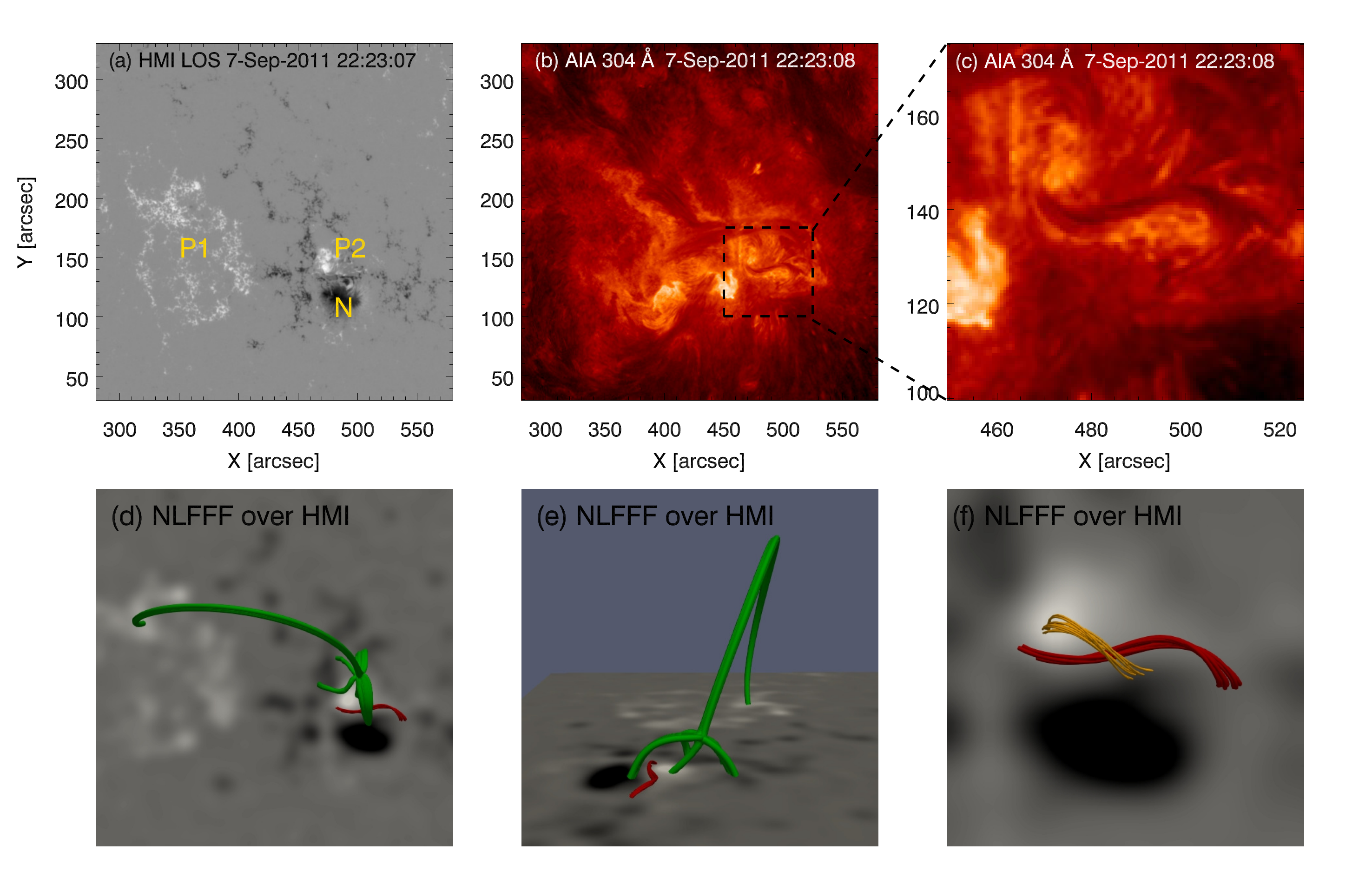}
                \caption{Magnetic modeling of the flare-hosting AR and its comparison with EUV imaging observations. Upper: HMI LOS magnetogram (a) and AIA 304~{\AA} images (b--c) of   pre-eruption AR\@. Panel c gives a zoomed-in view of the region outlined by the dashed box in panel b. Bottom: Results of the NLFFF extrapolation seen from different perspectives. The fan-spine structure, MFR field lines, and overlying arcades are colored green, red, and orange, respectively. On the bottom boundary, an HMI $B_z$ magnetogram in CEA projection is displayed as the background with a linear scale saturating at $\pm$1000~G. }
                \label{fig3}
        \end{figure*}

        The appearance of multiple flare ribbons and different sets of flare loops in this event implies a complex magnetic configuration of the AR, in which multiple magnetic reconnections may be involved. Figure \ref{fig3}a shows a line-of-sight (LOS) magnetogram of the pre-eruption AR obtained by the Helioseismic and Magnetic Imager \citep[HMI,][]{schou2012} on board \emph{SDO}. In addition to a large-scale dipole consisting of  a pair of opposite polarities P1 and N, there is another parasitic positive polarity P2 embedded in the leading negative polarity N. By inspecting the simultaneous AIA 304~{\AA} image shown in Fig. \ref{fig3}b, it is found that the filament whose eruption drives the flare is initially located along the southern polarity inverse line (PIL) between polarities P2 and N. A zoomed-in view  clearly reveals a forward \textsf{S} shape for this filament (Fig. \ref{fig3}c). 
        
        To further explore the coronal magnetic topology of the AR, we performed a nonlinear force-free-field (NLFFF) extrapolation based on the photospheric vector magnetogram with HMI\@. To prepare the boundary condition, we remapped the original HMI magnetic vector in a cylindrical equal-area (CEA) projection \citep[cf.][]{gary1990}, and applied a pre-processing procedure to minimize the net magnetic force and torque on the bottom boundary \citep{wiegelmann2006}. Using the optimization method \citep{wheatland2000,wiegelmann2004}, the NLFFF extrapolation is carried out in a Cartesian box of $424\times263\times263$~Mm$^{3}$.
        
        Figures \ref{fig3}d--f demonstrate the results of the NLFFF extrapolation from different perspectives. A large-scale fan-spine magnetic structure is revealed, whose field lines surround a three-dimensional magnetic null point lying 14.6 Mm above the parasitic polarity P2 (green lines in Figs. \ref{fig3}d and e). This  fan-spine topology facilitates the appearance of a quasi-circular flare ribbon at the fan skirt, as seen in this event.  Under the fan dome, twisted field lines are found to lie along the PIL where the above-mentioned sigmoid filament is located (red lines in Figs. \ref{fig3}d and e). In a zoomed-in view (Fig. \ref{fig3}f), these field lines exhibit nearly the same forward \textsf{S}-shaped morphology as the filament, and their twist number is estimated to be over 1.5 turns. In addition, the overlying envelope field lines are right-skewed (orange lines in Fig. \ref{fig3}f).  All these facts indicate that the twisted field lines correspond to a positive helicity MFR, which hosts a sinistral filament \citep{martin1992,chen2014}. Hereafter we  use the terms MFR and filament interchangeably, even though the filament may just occupy a small fraction of the whole MFR system. In passing, we note that this northern hemisphere filament does not obey the well-known hemispheric preference that the chirality of a filament is predominately dextral (sinistral) in the northern (southern) hemisphere \citep[e.g.,][]{martin1994,ouyang2017}, and is therefore   more prone to a major eruption \citep{bao2001}.
                
        The magnetic topology of the AR has been explored in many previous studies \citep[e.g.,][]{zhang2015,janvier2016,dai2018a,jiang2018}. For the 2011 September 6 X2.1-class flare occurring one day before from the same AR, the magnetic modeling also reveals a fan-spine structure, but the outer spine therein extends northwestward to another polarity outside the AR rather than northeastward to polarity P1 in this event \citep{janvier2016,jiang2018}. Probing the detailed magnetic process that causes the transition of outer spine orientation between these two X-class flares is beyond the scope of this work. Nevertheless, as a result of this transition, more magnetic flux could be accumulated directly above the AR, which would be effectively disturbed by the eruption of the MFR from below the fan dome.
        
        \subsection{Nonradial eruption of the MFR}\label{sec32}
\begin{figure*}
                \centering
                \includegraphics[width=14cm]{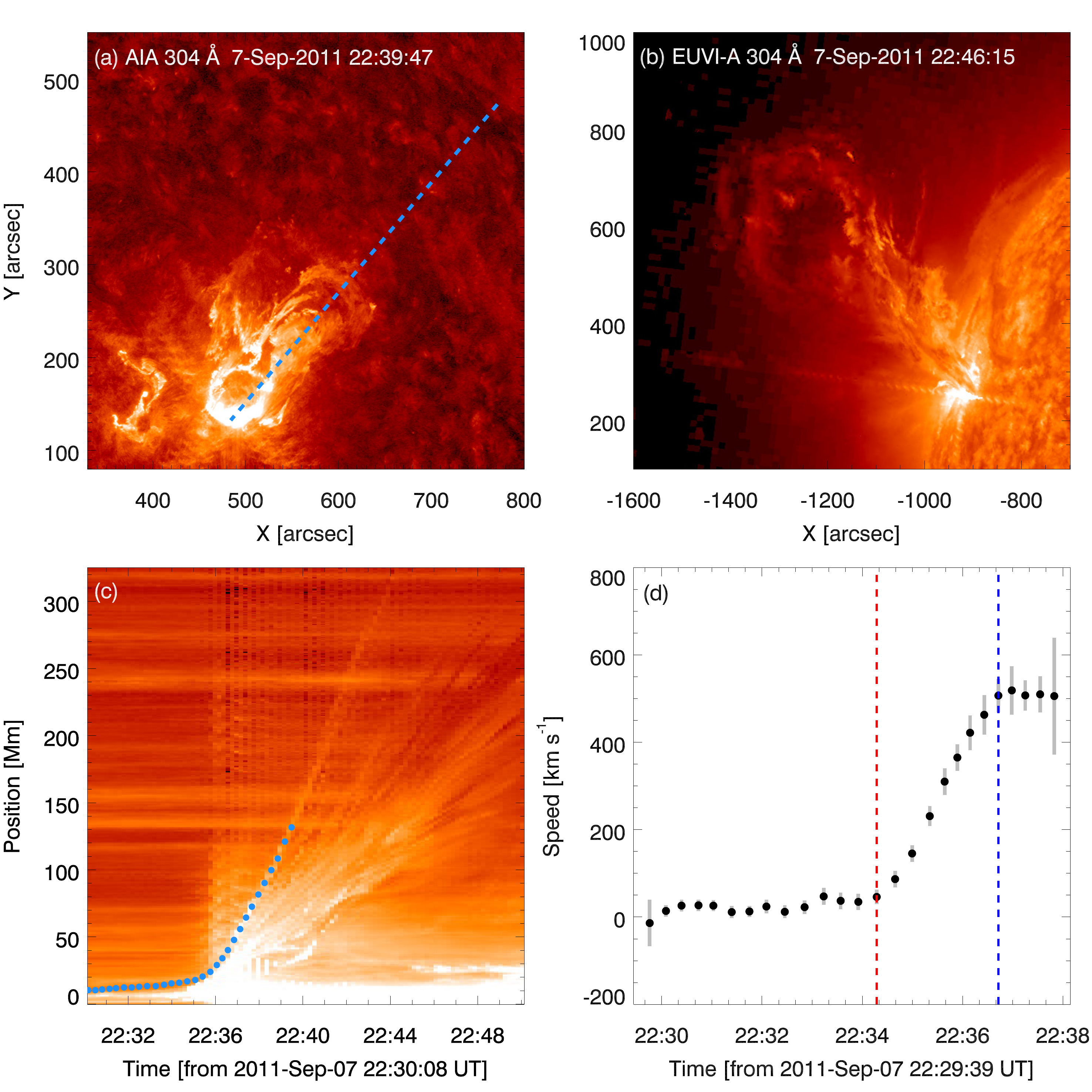}
                \caption{Eruption process of the flare-driving filament. Panels a and b demonstrate the snapshots of the filament eruption seen in \emph{SDO}/AIA 304 {\AA} and \emph{STEREO-A}/EUVI 304 {\AA}, respectively. Panel c shows the time-distance stack plot of the AIA 304~{\AA} intensities along the slice drawn in panel a. Panel d displays the velocity evolution of the filament leading edge along the selected slice, which is traced by the filled circles plotted in panel c. The vertical dashed lines in panel d divide the eruption process into three stages.}
                \label{fig4}
        \end{figure*}

        Online Animation 2 demonstrates the eruption process of the sigmoid filament seen in AIA 304~{\AA}, with one snapshot at the  fully developed stage of the filament eruption displayed in Fig. \ref{fig4}a. When viewed from the Earth's perspective, the filament erupts in a northwest direction, along which we selected a slice passing across the apex of the erupting filament  (blue dashed line in Fig. \ref{fig4}a) to trace its kinematic evolution. Figure \ref{fig4}c shows the time--distance stack plot of the AIA 304 {\AA} intensity along the slice. In the stack plot, the  filament first exhibits a slow rise, then experiences an impulsive acceleration, and quickly turns to a fast propagation. During the drastic acceleration, the originally dark filament materials are brightened up by the flare heating, and its body splits apart into several threads, which separate from each other during the following eruption. It is noted that some filament segments finally fall down and impact the solar surface, causing local brightenings. As shown in online Animation 2, there are two episodes of the falling filament impact that occur at a relatively nearby location around $(X,Y)=(580\arcsec, 320\arcsec)$ during 23:05--23:55 UT, and a farther one around $(X,Y)=(460\arcsec, 540\arcsec)$ later during 00:15--00:35 UT of the next day.
        
        Through visual inspection, we then identified the leading edge of the erupting filament, whose displacement against the plane of the sky (POS) is tracked by the filled circles plotted in Fig. \ref{fig4}{c}. To minimize subjective bias introduced by the manual inspection, at each point we repeated the position measurement five times, and adopted the average over these measurements as the measured value, and used the corresponding standard deviation as the measured uncertainty. Based on the time--distance information, we derived the POS velocity of the filament leading edge, whose evolution is plotted in Fig. \ref{fig4}d. The three-stage eruption process of the filament is now quantitatively established. During the slow rise stage, the speed of the filament is only a few tens of km~s$^{-1}$. Within a short period of 22:34:20--22:36:40~UT, the filament impulsively accelerates from below 50~km~s$^{-1}$ to over 500~km~s$^{-1}$. After that, the filament moves outward at a fast but relatively constant speed of about 500~km~s$^{-1}$.
        
\begin{figure*}
                \centering
                \includegraphics[width=14cm]{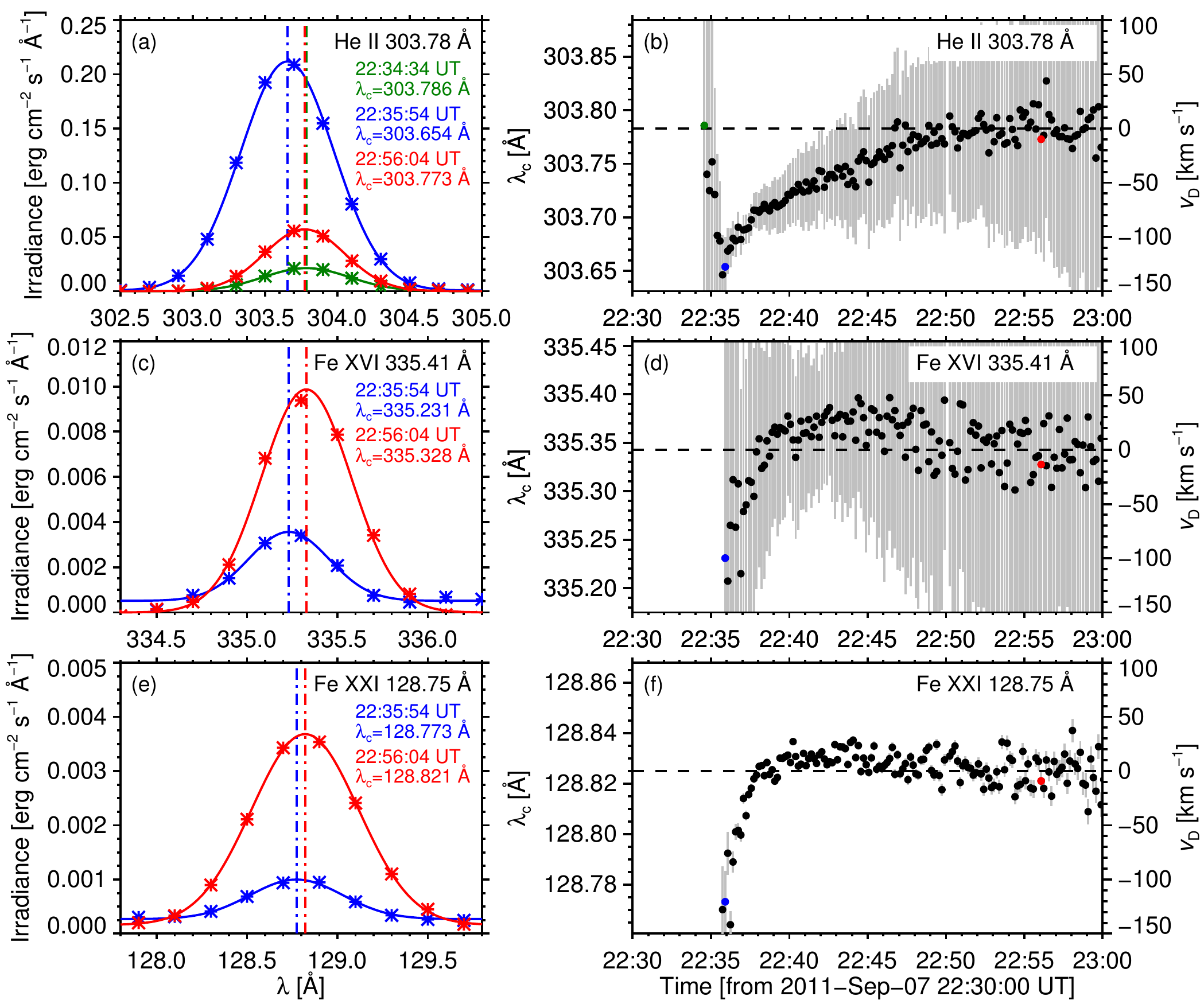}
                \caption{Results of the spectral fitting to EVE spectra and their temporal variations. Left: Background-subtracted spectra of three EVE spectral lines (asterisks) observed at different stages of the flare (see legend for color-coding) as well as the results of a single-Gaussian fit (solid lines) to the observed spectra. The vertical dash-dotted lines denote the fitted line center. Right: Temporal variations in the line center wavelength for the three EVE lines based on the Gaussian fit, with the fitting results plotted as filled circles and the corresponding fitting errors as gray error bars. In each panel the colored circles represent the same fitting results for the corresponding spectral line displayed in the left, and the horizontal dashed line indicates the rest wavelength of the line center, which is derived by averaging the fitting results over an interval of 22:50--23:00 UT when the main flare has turned to a gradual decay. With this reference wavelength, the variation in line center wavelength (scaled by the left axis) is further converted to an equivalent variation of Doppler velocity (scaled by the right axis).}
                \label{fig5}
        \end{figure*}

        In addition to the POS velocity, we also estimated the LOS velocity of the erupting filament based on spectral fitting to the EVE spectra. Due to the optical design of MEGS-A, any off-axis light across the 0.5$^{\circ}$ field of view (FOV) of the Sun (e.g., that from the limb) causes a position-dependent wavelength offset with respect to the light from the disk center, which makes interpreting the Doppler shifts derived from full-Sun EVE spectra complicated \citep[cf.][]{chamberlin2016}. Here we used the background-subtracted spectra to isolate flare-excess irradiance, which predominantly originates from a localized flaring region, and is therefore just subject to an approximately uniform wavelength shift. Figure \ref{fig5}a shows the background-subtracted spectra of the \ion{He}{II} 303.78~{\AA} line observed at three different moments. The spectra can be well fitted using a single-Gaussian function, with the fitting results overplotted in the figure. Compared with the other times, an obvious blueshift is revealed from the fitted line center at 22:35:54~UT (blue), a moment when the erupting filament is experiencing the  impulsive POS acceleration (see Fig. \ref{fig4}d). Since the flare-excess emissions at the early stage are dominated by the erupting filament (see Animation 2), the blueshift should reflect a LOS eruption of the filament away from the solar disk.   
        
        We fitted all the flare-excess spectra of the \ion{He}{II} 303.78~{\AA} line that have high enough line intensities; a threshold of  3\% increase above the background level was adopted for \ion{He}{II} 303.78~{\AA}. The temporal variation of the fitted line center wavelength is plotted in Fig. \ref{fig5}b. As shown in the figure, the line center first experiences a rapid blueward shift, indicating an impulsive acceleration of the erupting filament along the LOS\@. The filament LOS acceleration inferred from the EVE spectra starts nearly simultaneously with the POS acceleration, but stops about one minute earlier, as indicated by the time evolution of the Doppler shift for the line center. Nevertheless, we do not attribute the diminishing blueshift to a deceleration of the filament, but to a transition of the main emission contributor from the erupting filament to flare ribbons, in the latter of which the cold \ion{He}{II} 303.78~{\AA} line should be somewhat redshifted due to chromospheric condensation. In addition, the eruption of the filament itself in a direction of increasing longitude, as revealed from the AIA images, should further impose a redshifted instrumental offset. Due to these limitations, the spatially unresolved EVE observations cannot capture the full acceleration process of the filament as revealed in AIA\@. As the main flare turns to a gradual decay, the fitted line center wavelength finally recovers to a relatively stable level. As done in \citet{hudson2011a}, we selected an interval of 22:50--23:00~UT during the gradual decay, and took the average of the fitting results over this interval  as the  ``rest''  wavelength of the line center. With this reference wavelength, a maximum blueshifted Doppler (LOS) velocity of 135 km~s$^{-1}$ at 22:35:44~UT is derived for the filament eruption.
        
        We also fitted the spectra of the \ion{Fe}{XVI} 335.41~{\AA} (Figs. \ref{fig5}c and d) and \ion{Fe}{XXI} 128.75 {\AA} (Figs. \ref{fig5}e and f) lines, which are sensitive to warm and hot plasmas, respectively. Subject to low line intensities in the early stage, the impulsive acceleration of the filament cannot be captured in these two lines. However, similar maximum Doppler velocities of 120--140 km~s$^{-1}$ are revealed, which occur almost simultaneously with the maximum Doppler velocity in \ion{He}{II} 303.78~{\AA}  (Figs.~\ref{fig5}d and f). The consistent Doppler velocities appearing in the different spectral lines reflect a true LOS velocity of the erupting filament, and meanwhile imply a multithermal nature of the MFR system \citep{wang2022}. The blueshifts diminish faster in \ion{Fe}{XVI} 335.41~{\AA}  and \ion{Fe}{XXI} 128.75~{\AA}, possibly because the contributor of the flare emission changes faster from the erupting filament to the flare loops and the flare ribbons in these passbands.
        
        If we assume that the complete LOS acceleration of the filament is synchronous with the POS acceleration, the final LOS velocity at the end of the filament acceleration is estimated to be 250~km~s$^{-1}$, which is only half of the final POS velocity. Considering that the AR location of N14W31 is not far from the disk center, the eruption of the MFR should follow a nonradial  direction $\sim$30$^{\circ}$   inclined toward the solar surface. Figure \ref{fig4}b displays an image of the filament eruption in 304~{\AA} taken by the EUV Imager \citep[EUVI,][]{wuelser2004} on board the \emph{Ahead} spacecraft of the \emph{Solar Terrestrial Relations Observatory} \citep[\emph{STEREO-A},][]{kaiser2008}. At the time of the flare, \emph{STEREO-A}  is orbiting 103$^{\circ}$ ahead of the Earth, which therefore permits an edge-on view of the flare evolution. As expected, the erupting filament seen from the \emph{STEREO-A} perspective  shows an obvious inclination toward the solar north pole.
        
        \section{Multiple reconnections driven by the erupting MFR}\label{sec4}
\begin{figure*}
                \centering
                \includegraphics[width=12cm]{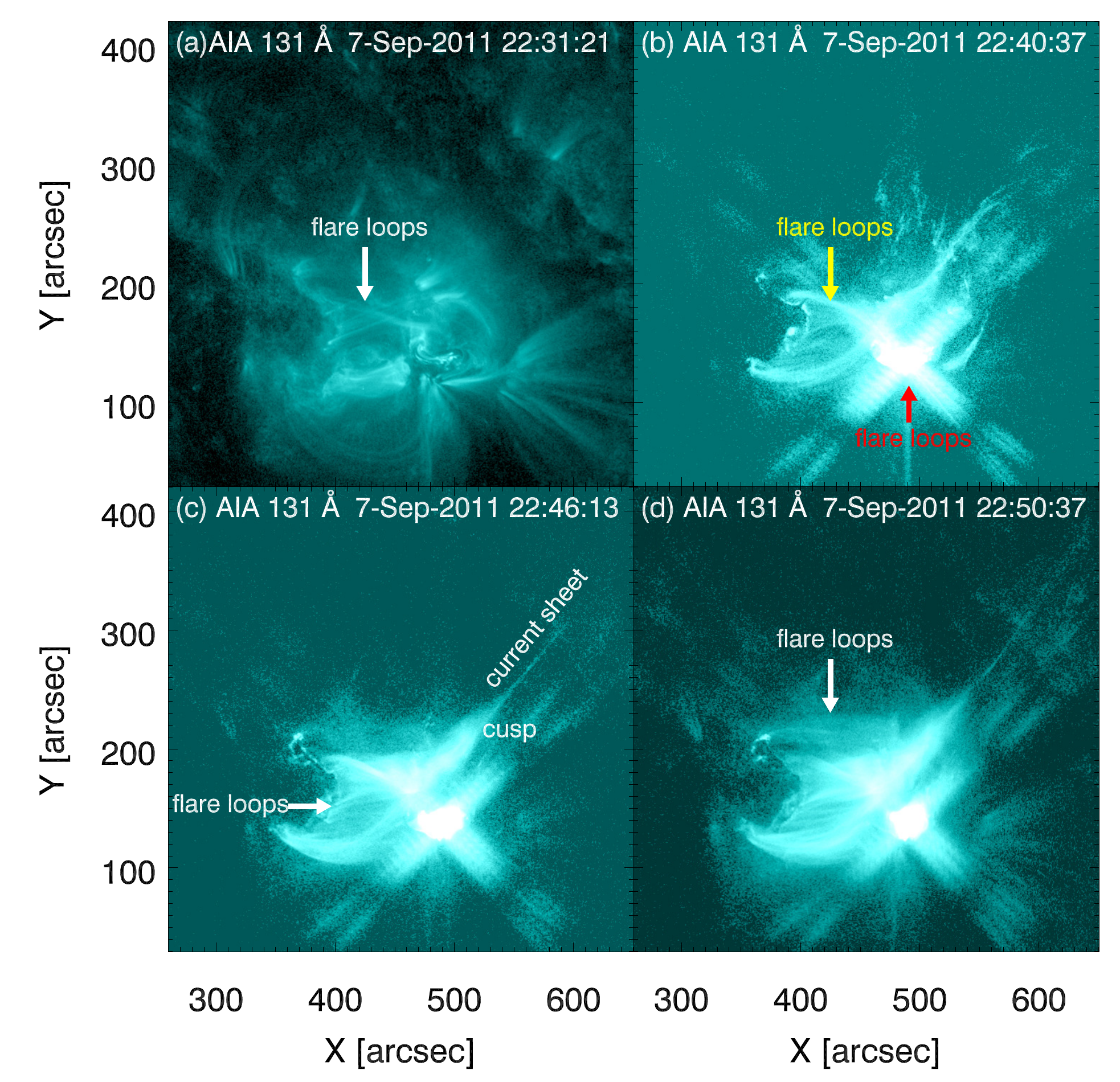}
                \caption{Multiple MFR-driven magnetic reconnections as revealed in AIA 131 {\AA}. The arrows and notations highlight some characteristic flare loops and structures (see   text for details).} 
                \label{fig6}
        \end{figure*}

        Under the complex magnetic configuration, multiple magnetic reconnections are invoked during the whole process of the MFR eruption.  Some snapshots of the individual reconnections, as illustrated by newly formed flare loops seen in the hot  AIA 131 {\AA} passband, are extracted from online Animation 3 and presented in Fig. \ref{fig6}. Accompanying the slow rise of the MFR (22:30--22:34 UT),  slight brightening is first seen in loops extending northeastward from above the MFR (indicated  by the arrow in Fig. \ref{fig6}a).  The position and orientation of the brightening loops are generally consistent with the outer spine lines revealed in the NLFFF extrapolation (see Fig. \ref{fig3}d), except that the extrapolated outer spine extends $\sim$30~Mm farther. We attribute this brightening to a mild magnetic reconnection taking place around the null point, which is believed to be triggered by the destabilization of the underlying MFR\@. 
        
        It is also interesting to note the presence of a pre-event brightening during the preceding C-class flare (22:13--22:30 UT), which occurs below the outer spine and adjacent to the fan dome (see online Animation 3). The magnetic reconnection responsible for this phase may remove some outside magnetic field lines initially inclined on the fan dome, partially relax the magnetic confinement over the embedded MFR, and possibly trigger some kind of MHD instability (e.g., torus instability; \citealt{Kliem2006}) for the MFR\@. As a consequence, faint brightening signatures have been observed in the above outer spine during this pre-event phase.
        
        The MFR-pushed null-point reconnection belongs to a ``breakout'' reconnection in nature \citep{antiochos1999}, whose effect is to facilitate further acceleration and eruption of the MFR by reducing the magnetic confinement above it \citep[e.g.,][]{sun2013}. The ensuing impulsive acceleration of the MFR in turn intensifies the null-point reconnection above, giving rise to a more prominent outer spine than before (highlighted by the yellow arrow in Fig. \ref{fig6}b). Beneath the MFR, meanwhile, a much more energetic flare reconnection sets in between the legs of field lines stretched out by the fast ascending MFR, acting like a standard flare reconnection in the CSHKP model. Since the MFR-stretched field lines at this stage are initially embedded under the fan dome, the resultant main flare loops are relatively short and compact (indicated by the red arrow in Fig. \ref{fig6}b).
        
        As the MFR breaks through the fan dome and rapidly propagates outward, it continues to stretch the overlying field lines outside the fan dome. As shown in Fig. \ref{fig6}c, a long straight current sheet is formed, which extends toward the northwest, consistent with the eruption direction of the MFR\@. At the lower end, the current sheet is connected with the tip of a large-scale cusp structure, indicating a CSHKP magnetic reconnection analogous to the main flare reconnection, but on a much larger scale that produces significantly longer flare loops connected to the remote flare ribbon Rr (indicated  by the arrow in Fig. \ref{fig6}c).
        
        It is interesting to note that the orientation of the long flare loops produced by the large-scale CSHKP reconnection is roughly perpendicular to that of the main flare loops (see Fig. \ref{fig6}c). It implies that the reconnection-driving MFR may experience a clockwise rotation during its eruption, although the signature of rotation is largely blurred by the nonradial eruption and the splitting of the filament materials. On the one hand, the sense of the MFR rotation inferred from the flare loop orientation, follows the chirality rule (i.e., clockwise and counterclockwise respectively for a sinistral and dextral filament;   \citealt{zhou2020}). On the other hand, the rotation makes the axis of the MFR more perpendicular to the overlying arcades, which facilitates more effective stretching of the arcade field lines, consequently leading to a relatively strong reconnection despite a rather high reconnection site. 
        
\begin{figure*}
                \centering
                \includegraphics[width=14cm]{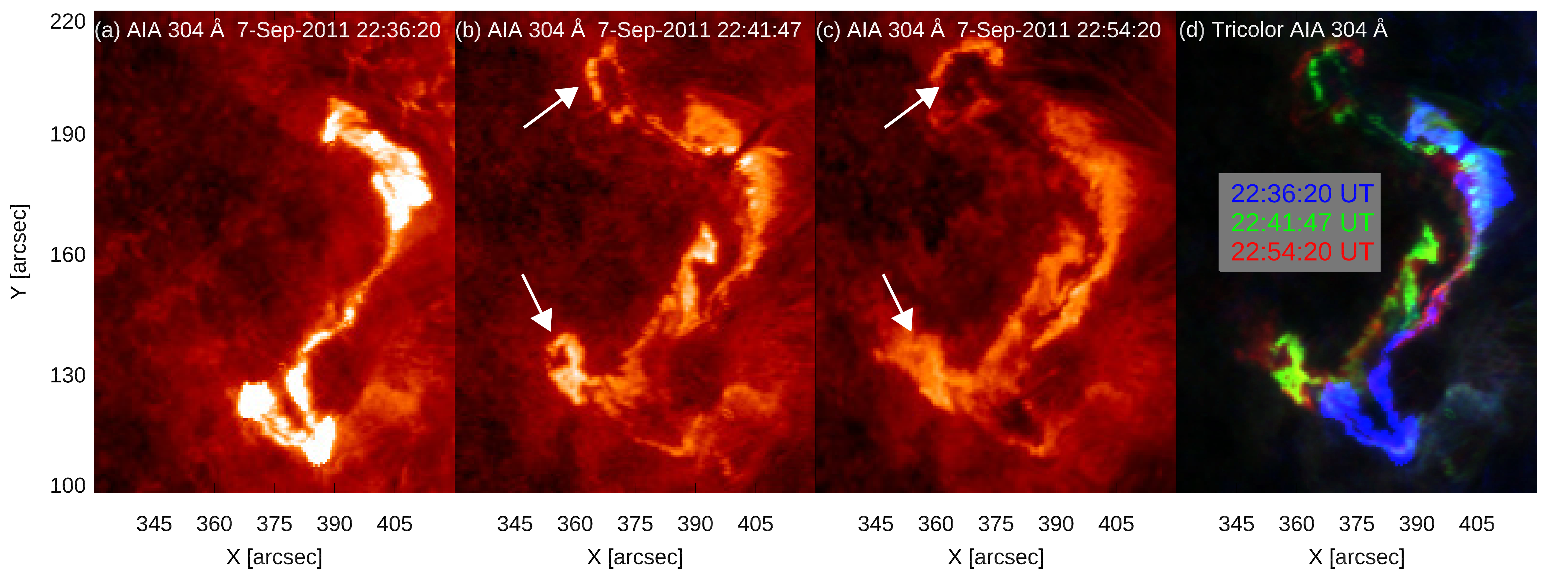}
                \caption{Dynamic evolution of the remote ribbon Rr seen in AIA 304 {\AA} (a--c) and the corresponding tricolor composite image (d). The arrows point to the northern and southern parts of the flare ribbon, which exhibit significant changes during the evolution.}
                \label{fig7}
        \end{figure*}

        The interaction of the MFR and the overlying field lines also squeezes the QSL between them, enhancing the electric current and inducing an additional magnetic reconnection there, known as QSL reconnection. This QSL reconnection is thought to take place in a halo surrounding the outer spine, therefore contributing to the formation and dynamic evolution of the remote flare ribbon Rr (located in polarity P1). Figure \ref{fig7} presents the morphological evolution of ribbon Rr seen in the chromospheric AIA 304 {\AA} passband. When it  appears, the ribbon Rr shows up with an elongated shape that extends 70 Mm in the north--south direction, suggesting a large-scale QSL-halo structure involved in the reconnection. Afterward, the ribbon Rr undergoes a dynamic evolution; the northern part of the ribbon moves mainly along the ribbon and extends to a farther distance, while the southern part drifts predominately perpendicular to the ribbon. Within a short interval of five minutes (22:36:20--22:41:47 UT) both parts are detached from the main body of the flare ribbon.  Meanwhile, the resultant flare loops are connected to where the northern and southern parts of Rr are initially located. In the south these flare loops  largely overlap with those produced by the large-scale CSHKP reconnection, while in the north, they are less affected by foreground contaminations and are thus more distinguishable, as seen in AIA 131 {\AA} (indicated by the arrow in Fig. \ref{fig6}d). It is worth mentioning that in the case of a confined event, most of the kinetic energy of an ``erupted-but-failed'' MFR may be finally dissipated through such a QSL reconnection, thus facilitating a considerably strong EUV late phase, as presented in \citet{liu2015}.
        
        Although spatially separated, the sequential magnetic reconnections are closely linked by the fast nonradial eruption of the MFR, with the transition from one reconnection to another happening in a short interval. Except the main flare reconnection, all other reconnections involve the production of long late-phase loops that make contributions to the late-phase emission.
        
        In passing, we note that the erupting MFR itself may also reconnect directly with ambient magnetic fields of different orientations, by which parts of filament materials are transported to the newly reconnected field lines, follow different trajectories, and finally fall down on the surface to produce the above-mentioned remote brightenings. This is largely consistent with the scenario recently proposed by \citet{Li2023}, and suggests a possible association between the nonradial MFR propagation and the partial filament eruption \citep{gilbert2007}. However, because of the spatial inconsistency, this type of magnetic reconnection is unlikely to be relevant to the production of the late-phase loops.
        
        \section{Long-lasting cooling of the late-phase loops} \label{sec5}
\begin{figure*}
                \centering
                \includegraphics[width=9cm]{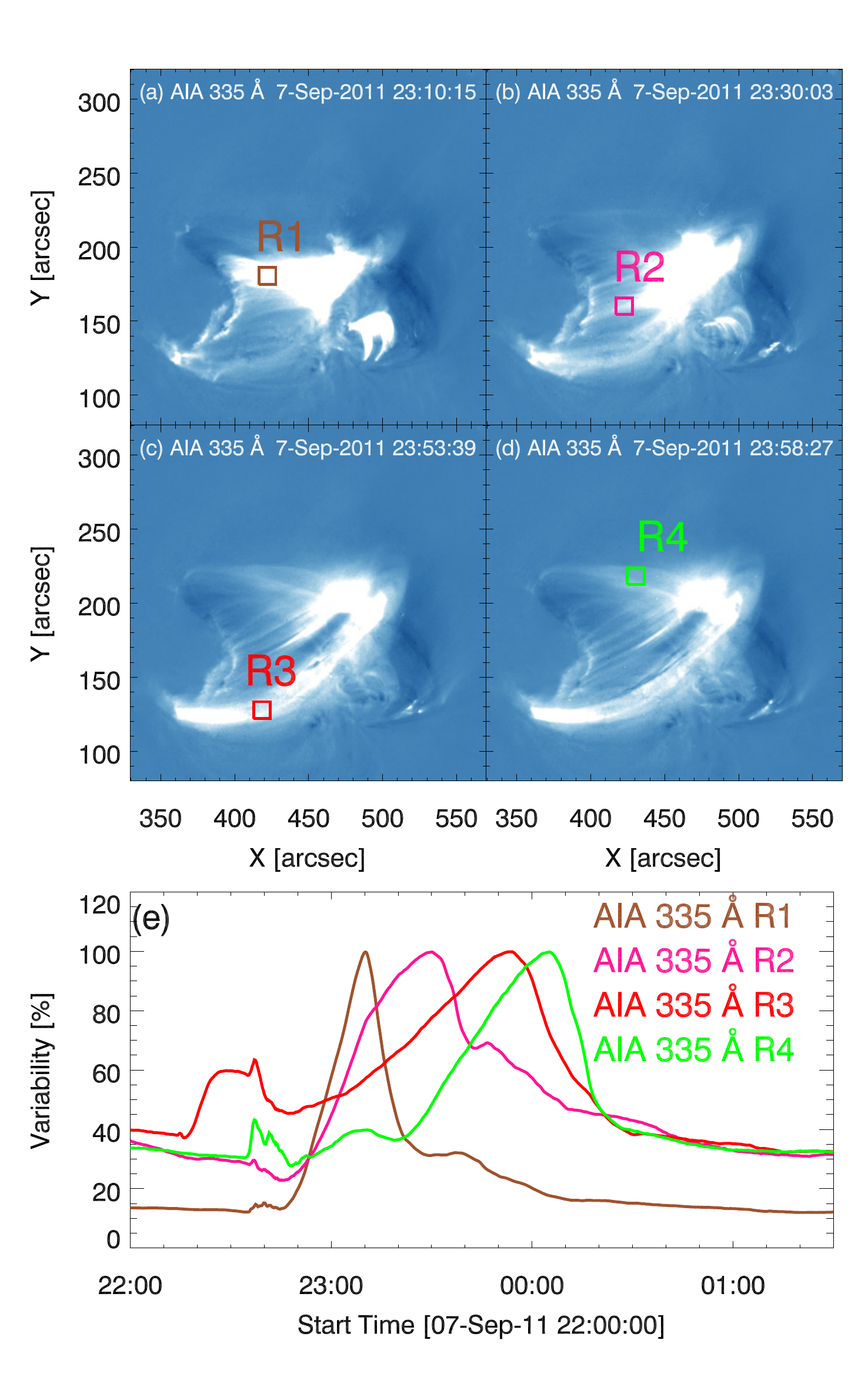}
                \caption{Morphology (a--d) and intensity variability (e) of   late-phase loops observed in AIA 335 {\AA}. The small boxes (labeled   R1--R4) outline four typical sets of late-phase loops, over which the light curves are plotted in panel e.} 
                \label{fig8}
        \end{figure*}

        Figures \ref{fig8}a--d show the morphology of the late-phase loops observed in AIA 335 {\AA}. To enhance brightening signatures, we use the base difference method (i.e., subtract a pre-event image taken around 22:00 UT from each frame of image. At different stages of the late-phase plateau, the emission originates from different late-phase loops. We identify four typical groups of late-phase loops with increasing lengths, and take four small boxes (labeled   R1--R4) overlaid on them as their proxies. By a comparison with Fig. \ref{fig6}, it is found that all the late-phase loops have high-temperature counterparts initially appearing in AIA 131 {\AA}: loop R1 corresponds to the outer spine lines energized by the null-point reconnection, loops R2 and R3 are   shorter and longer flare loops produced by the large-scale CHSKP reconnection, and loop  R4 is related to the QSL reconnection. Such a good correspondence suggests a cooling process of the late-phase loops from an initially high temperature over 10~MK. 
        
        The time profiles of intensity variability in the four loop regions are plotted in Fig. \ref{fig8}e. As expected, the longer the loops, the later the occurrence of the emission peaks. In particular, the fading-off of one set of late-phase loops is followed by the brightening-up of another set of longer loops, thus making the overall late-phase emission maintain a relatively constant level. Furthermore, all the intensity profiles generally show a slow rise followed by a fast decay in AIA 335 {\AA}. According to \citet{dai2018b}, this emission pattern may serve as an indicator for a long-lasting cooling process in sufficiently heated late-phase loops, where the warm coronal emission peak will occur in the radiative cooling stage. In this case, the follow-up emission should be significantly affected by coronal condensation, hence exhibiting a fast decay compared to the relatively gradual rise.

\begin{figure*}
                \centering
                \includegraphics[width=12cm]{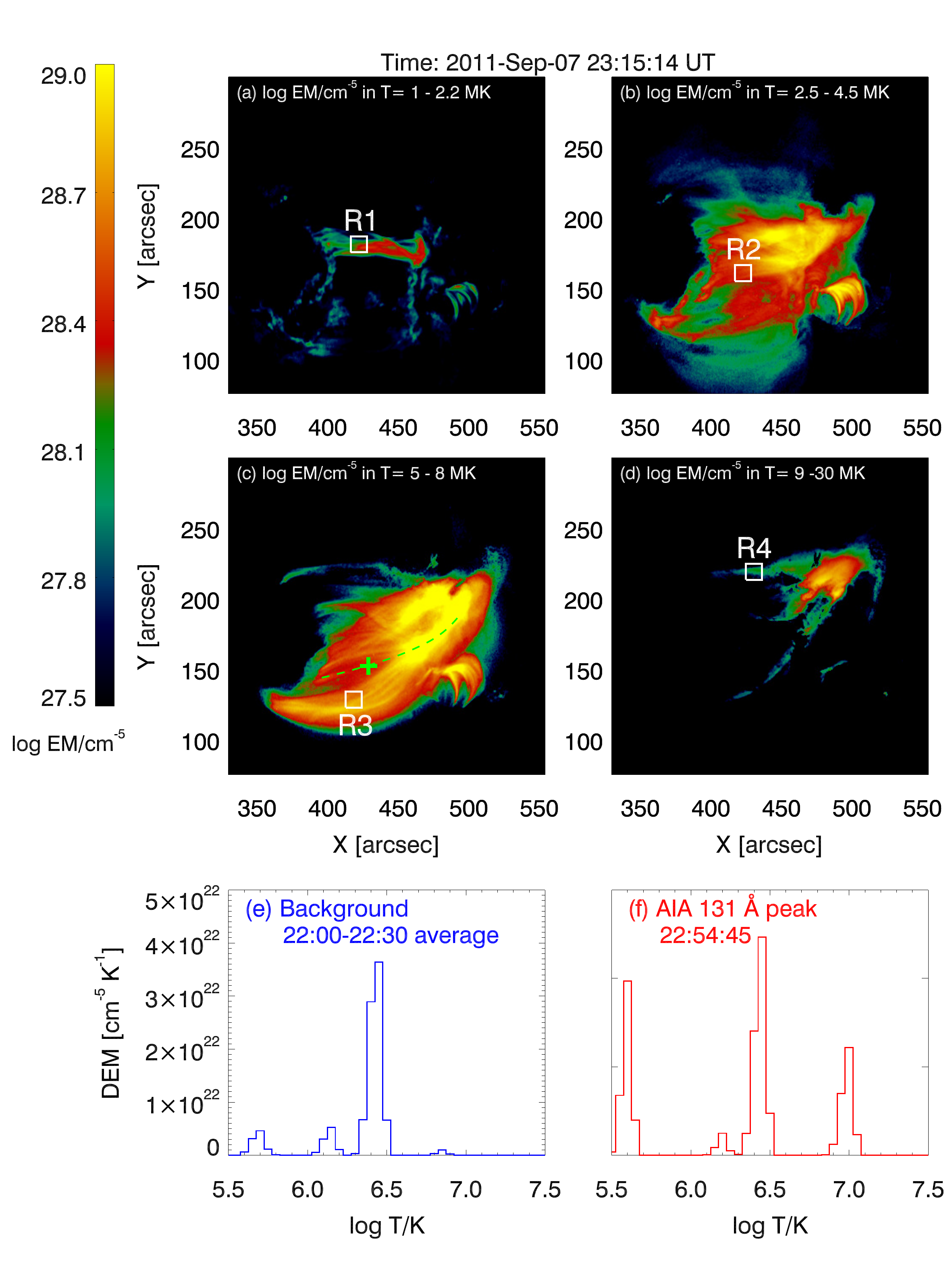}
                \caption{Results of DEM inversion to the AR\@. Panels a--d display the EM maps of the AR at 23:15:14 UT within four different temperature bins. The small boxes (labeled  R1--R4) are the same as in Fig. \ref{fig8}. Panels e and f plot the DEM distributions of a selected late-phase loop region (plus sign in the loop delineated by the dashed line) during the pre-event period and at the  corresponding AIA 131 {\AA} peak, respectively.}
                \label{fig9}
        \end{figure*}

        Using observations from the six coronal passbands of AIA, we further carried out differential emission measure (DEM) analysis to each pixel of the AR, from which the emission measure (EM) of late-phase loops over a certain temperature range can be constructed. Here we adopt the sparse algorithm \citep{cheung2015,su2018} to perform the DEM inversion. Figures \ref{fig9}a--d show EM maps of the AR at 23:15:14 UT in four different temperature bins, which reveal different temperature structures for the different late-phase loops:  R1 loops are predominantly seen in the cool 1--2.2 MK bin,  R2 and R3 loops present higher temperatures of 2.5--4.5 MK and 5--8 MK, respectively, while  R4 loops still keep a flaring plasma temperature of over 9 MK. This further validates a longer cooling process in longer late-phase loops. In passing, we note that the high-temperature cusp structure is successfully constructed in the EM map of 9--30 MK (Fig. \ref{fig9}d), which  reflects a sufficiently strong large-scale CSHKP reconnection in producing the late-phase loops.
        
        Finally, we use the results of DEM inversion to estimate the cooling time of a late-phase loop. From visual inspection, we select a late-phase loop (outlined by the dashed line in Fig. \ref{fig9}c) that is less affected by the background and foreground contaminations from other loops along the LOS, and pick up a small region (plus sign in Fig. \ref{fig9}c) overlapping the loop. The intensity profiles of this small region peak at 22:54:45 UT and 23:49:00 UT in the hot AIA 131 {\AA} and cool AIA 171 {\AA} passbands, respectively. Following \citet{dai2018b}, we take the time interval between these two peaks ($\sim$54 minutes) as the observed cooling time of this specific late-phase loop.
        
        Meanwhile, we plot in Figs. \ref{fig9}e and f the DEM distributions of the selected loop region during the pre-event period (22:00--23:30 UT) and at the AIA 131 {\AA} peak time, respectively. Compared with the quiescent level, a prominent bump at around $\log T\sim7.0$ is revealed in the DEM distribution at the AIA 131 {\AA} peak time. By isolating this hot component and assuming a LOS depth of the loop of 1.7 Mm (equivalent to the loop diameter), we drive a DEM-weighted temperature of $T\sim$10 MK and a number density of $n_e\sim10^{10}$~cm$^{-3}$ at the beginning of the loop cooling. In addition, we trace the loop spine based on a semicircular loop geometry, which yields a half-length of $L\sim60$ Mm for the late-phase loop. Using these loop parameters, we estimate the cooling time of the loop based on the analytical formula  
        \begin{equation}
                \tau _{\mathrm{cool}} \approx 2.35 \times 10^{-2} \frac{L^{5/6}}{T^{1/6}n_{\mathrm{e}}^{1/6}}
        \end{equation}
        given by \citet{cargill1995}. The theoretically predicted cooling time of the loop is $\sim$59 minutes, which is in good agreement with the observed value.
        
        \section{Summary and conclusions} \label{sec6}
        Using observations mainly with the instruments on board \emph{SDO}, we analyzed an X1.8-class solar flare occurring on 2011 September 7 from the complex AR 11283. For convenience, in Table \ref{tab:tab1} we summarize the timeline of the activities involved in or associated with the event.
        
        The EVE observations reveal a nearly constant plateau of the warm coronal emission following the main flare peak, which lasts for almost one hour before falling back to the background level. Meanwhile, the AIA observations clearly show that the plateau-like warm coronal emission originates from a significantly larger region outside the main flare region, validating the identification of an EUV late phase for this emission plateau.  
        
        The magnetic modeling of the AR reveals a filament-hosting MFR embedded under a fan-spine topology, whose destabilization and eruption drive multiple magnetic reconnections that power the atypical EUV late-phase flare. The special dynamics of the erupting MFR plays an important role in initiating the multiple reconnections. First, by tracking the kinematics of the MFR eruption both in the POS and along the LOS, we infer a nonradial eruption of the MFR. Along the eruption direction, the MFR may squeeze the magnetic null point efficiently,  therefore intensifying the null-point reconnection therein. Second, the erupting MFR is believed to experience a clockwise rotation in a pattern following the chirality rule \citep{zhou2020}. When the MFR breaks through the fan dome, its axis moves closer to perpendicular to the overlying large-scale arcades, which facilitates a more effective stretching of the arcade field lines, consequently leading to a sufficiently strong large-scale CSHKP reconnection. Finally, the interaction of the MFR and the overlying field lines results in a QSL reconnection, as revealed from a dynamic evolution of the remote flare ribbon, which makes an additional contribution to the production of late-phase loops. These effective reconnections ensure a sufficiently strong late-phase emission. We note that the absolute level of the late-phase peak in this flare is the highest among all the late-phase flares from the same AR \citep[cf.][]{zhong2021}. 
        
        Because of the fast propagation of the MFR, all the magnetic reconnections take place in a short interval. In this sense, the late-phase loops are heated nearly simultaneously with the main flare loops. The morphology of the late-phase loops can be sequentially traced in AIA passbands with decreasing response temperatures (in a sequence from hot AIA 131 {\AA}, to warm AIA 335 {\AA}, and eventually to cool AIA 171 {\AA}), suggesting a cooling process of the late-phase loops from an initially high temperature of over 10 MK \citep{liu2015,dai2018a}.
        
        Nevertheless, the late-phase loops produced by the multiple reconnections possess different lengths. Since the cooling time of a loop is largely proportional to its half-length \citep{cargill1995}, these different groups of late-phase loops will cool down at different  rates, which makes their warm coronal emissions peak at different time instants. If the peak intensities of these late-phase loops are of similar magnitudes, the overall  emission summed over all these late-phase loops will exhibit an elongated plateau, as is seen in this event. By comparison, in an ordinary EUV late-phase flare the late-phase emission is contributed by late-phase loops of similar lengths, and thus the emission typically presents a well-resolved late-phase peak.
        
        Qualitatively, the AIA 335 {\AA} light curves of the individual late-phase loops show a slow rise followed by a fast decay. This emission pattern conforms to the long-lasting cooling scenario as proposed by \citet{dai2018b}. Quantitatively, we estimate the cooling time of a specific late-phase loop based on the loop properties derived from the DEM inversion. The theoretically predicted cooling time is in good agreement with the observed value.
        
        To summarize, in this atypical EUV late-phase flare, the plateau-like late-phase emission originates from the long-lasting cooling of different groups of late-phase loops, which are produced in multiple magnetic reconnections (including a null-point reconnection, a large-scale CHSKP reconnection, and a QSL reconnection) driven by the nonradial eruption of a MFR.        
        
        
        \begin{acknowledgements}
                We are grateful to the anonymous referee whose valuable comments and suggestions led to a significant improvement of the manuscript. This work was supported by National Natural Science Foundation of China under grants 11733003 and 12127901. Y.D. is also sponsored by National Key R\&D Program of China under grants 2019YFA0706601 and 2020YFC2201201, as well as Frontier Scientific Research Program of Deep Space Exploration Laboratory under grant 2022--QYKYJH--HXYF--015. \emph{SDO} is a mission of NASA's Living With a Star (LWS) program.
        \end{acknowledgements}
        

        \clearpage

        \clearpage
        
        \clearpage
        
        \clearpage
        
        \clearpage
        
        \clearpage
        
        \clearpage
        
        \clearpage

        \clearpage
        
        \begin{table}[htbp]
                \centering
                \caption{\centering Time history of the event}
                \begin{tabular}{ccc}
                        \toprule
                        \toprule
                        \textbf{Time (UT)} & \textbf{Activities} & \textbf{Instruments} \\
                        \midrule
                        22:13, 22:18, 22:26 & Start, peak, and end of the precursor C-class flare & \emph{GOES} \\
                        \midrule
                        \multirow{2}[2]{*}{22:13--22:30} &  Faint brightening of the outer spine, & \multirow{2}[2]{*}{\emph{SDO}/AIA} \\
                        &  C-class flare brightening below the outer spine &  \\
                        \midrule
                        22:32, 22:38, 22:44 & Start, peak, and end of the major X-class flare & \emph{GOES} \\
                        \midrule
                        22:37:04 & Peak of the \ion{He}{II} 304~{\AA} irradiance & \emph{SDO}/EVE \\
                        \midrule
                        22:40:24 & Peak of the \ion{Fe}{XX/XXIII} 133~{\AA} irradiance & \emph{SDO}/EVE \\
                        \midrule
                        22:41:54 & Peak of the \ion{Fe}{XVI} 335~{\AA} irradiance & \emph{SDO}/EVE \\
                        \midrule
                        22:30--22:34 & Slow rise of the filament, slight brightening of the outer spine & \emph{SDO}/AIA \\
                        \midrule
                        \multirow{4}[2]{*}{22:34--22:37} & Impulsive acceleration of the filament in a northwest direction,  & \multicolumn{1}{c}{\multirow{2}[1]{*}{\emph{SDO}/EVE, AIA,}} \\
                        & blueshifted Doppler (LOS) velocity, &  \\
                        & appearance of multiple flare ribbons (Rc, Ri, Rr),  & \multirow{2}[1]{*}{\emph{STEREO-A}/EUVI} \\
                        & more prominent outer-spine loops &  \\
                        \midrule
                        \multirow{3}[2]{*}{22:37--22:55} & Dynamic evolution of the remote ribbon Rr, & \multirow{3}[2]{*}{\emph{SDO}/AIA} \\
                        & long current sheet connected with a large-scale cusp structure, &  \\
                        & appearance of late-phase loops in hot passbands &  \\
                        \midrule
                        22:52--23:50 & Plateau-like warm coronal late-phase emission & \emph{SDO}/EVE, AIA \\
                        \midrule
                        23:05--23:55 &   Impact of falling filament material on the surface  $\sim(580\arcsec,320\arcsec)$ & \emph{SDO}/AIA \\
                        \midrule
                        00:15(+1)--00:35(+1) & Impact of falling material filament on the surface $\sim(460\arcsec,540\arcsec)$ & \emph{SDO}/AIA \\
                        \bottomrule
                \end{tabular}%
                \label{tab:tab1}%
        \end{table}%
        
\end{document}